# A Non-standard Model for Microbial Enhanced Oil Recovery Including the Oil-water Interfacial Area

D. Landa-Marbán, F.A. Radu and J.M. Nordbotten

**Abstract**

In this work we present a non-standard model for microbial enhanced oil recovery including the oil-water interfacial area. Including the interfacial area in the model, we eliminate the hysteresis in the capillary pressure relationship. One of the characteristics that a surfactant should have, it is biological production at the oil-water interface. Therefore, we consider the production rate of surfactants not only as a function of the nutrient concentration, but also the interfacial area. To solve the model equations, we use an efficient and robust linearization scheme that considers a linear approximation of the capillary pressure gradient. A comprehensive, 1D implementation based on two-point flux approximation of the model is achieved. We consider different parameterizations for the interfacial tension and residual oil saturation reduction. Illustrative numerical simulations are presented, where we study the spatial distribution and evolution in time of the average pressure, water saturation, interfacial area, capillary pressure, residual oil saturation and bacterial, nutrient and surfactant concentrations. Inclusion of the interfacial area in the model leads to different predictions of oil recovery. The model can also be used to design new experiments contributing to a better understanding and optimization of MEOR.

**Introduction**

Today, oil is one of the most valuable natural resources on Earth. Since we started to use petroleum in society, we have developed different techniques in order to recover the most oil possible from the reservoirs. Using and maintaining the initial reservoir pressure, it is possible to recover up to 50% oil (Patel et al. 2015). Then, enhanced oil recovery (EOR) techniques are used in order to improve the extraction.

Microbial enhanced oil recovery (MEOR) is an environmentally friendly and economically feasible EOR technique that uses the different bacteria products (acids, biomass, gases, polymers, solvents and surfactants) for improving the oil recovery. Among all advantages of using MEOR, we have that both bacteria and nutrients are easy to obtain and handle in the field, the bacterial activity increases over time and the field facilities just need minor modifications (Lazar et al. 2007, Patel et al. 2015). However, some of the challenges to overcome relating to MEOR field application are the transportation of all necessary components to the desired zones, minimize undesirable secondary activity and repair the loosing of components due to early absorption (Lazar et al. 2007). Notwithstanding these disadvantages, we are optimistic about an increase of the number of companies applying MEOR.

Most of the current models representing systems involving interfaces neglect the inclusion of the interfacial in the mathematical model. Measuring interfaces is difficult, expensive and occurs under simplified assumptions (Porter et al. 2010). Nevertheless, interfaces play a crucial role in the evolution of the physical quantities. The reason is that the bigger the contact area, the more influential the surface effects.

There exist different models for MEOR (Kim 2006, Nielsen et al. 2010a, Li et al. 2011); however, to our knowledge, there is no a model in the current literature that includes the oil-water interfacial area. The inclusion of the oil-water interface will allow to reduce the hysteresis in the capillary pressure (Hassanizadeh and Gray 1993) and to include the chemotaxis (unpubl. result Landa-Marbán, D., Radu, F.A., and Nordbotten, J.M. [2017] Modeling and simulation of microbial enhanced oil recovery including interfacial area. Version 1. arXiv:1612.04663v1 [physics.flu-dyn]).

The aim of this work is to present an accurate numerical simulator for two-phase flow and transport equations for bacteria, nutrients and surfactants, including the oil-water interfacial area and the surfactant effects.

**Theory**

Let us consider a porous medium filled with oil and water. The mass balance equation for the $\alpha$ phase ($\alpha = o$ for the oil and $\alpha = w$ for the water) is written as

$$\frac{\partial(\phi s_\alpha)}{\partial t} - \nabla \cdot \left(\frac{k_{r\alpha}}{\mu_\alpha} k (\nabla p_\alpha - g)\right) = \frac{F_\alpha}{\rho_\alpha} \tag{1}$$

where $\phi$ is the porosity, $s_\alpha$ is the saturation, $k_{r,\alpha}$ the relative permeability, $\mu_\alpha$ is the fluid viscosity, $k$ the absolute permeability, $p_\alpha$ is the pressure, $\rho_\alpha$ is the fluid density, $g$ is the gravitational acceleration vector and $F_\alpha$ is the source/sink term.

For the oil and water relative permeability curves, we use the Corey correlations (Lake 1989)

$$k_{ro} = k_{rowi} \left(\frac{1-s_w-s_{or}}{1-s_{wi}-s_{or}}\right)^b \tag{2}$$

$$k_{rw} = k_{rwor} \left(\frac{s_w-s_{wi}}{1-s_{wi}-s_{or}}\right)^a \tag{3}$$

where $k_{rowi}$ is the endpoint relative permeability for oil (at $s_{wi}$), $k_{rwor}$ is the endpoint relative permeability for water (at $1 - s_{or}$), $s_{or}$ is the residual oil saturation, $s_{wi}$ is the irreducible water saturation and $a$ and $b$ fitting parameters.

The porous medium is just filled with oil and water, then

$$s_o + s_w = 1. \tag{4}$$

On the core-scale, the capillary pressure is given by

$$p_o - p_w = p_c. \tag{5}$$

In standard models, the capillary pressure is given just as a function of the water saturation. However, the capillary pressure presents hysteresis (Nordbotten and Celia 2011). In Hassanizadeh and Gray (1993), they proposed that inclusion of the oil-water interfacial area in the capillary pressure relation reduces the hysteresis. In this work, we use the following interfacial area relation (Joekar-Niasar and Hassanizadeh 2012)

$$a_{ow}(s_w, p_c) = \alpha_1 s_w^{\alpha_2} (1-s_w)^{\alpha_3} p_c^{\alpha_4} \tag{6}$$

where $\alpha_1$, $\alpha_2$, $\alpha_3$ and $\alpha_4$ are fitting parameters. From the previous equations, it is possible to find $p_c(s_w, a_{ow})$.

Defining the average pressure $p = 0.5(p_o + p_w)$, the phase mobility $\lambda_\alpha = k_{r,\alpha}/\mu_\alpha$, the sum of phase mobility $\lambda_\Sigma = \lambda_o + \lambda_w$ and the subtraction of phase mobility $\lambda_\Delta = \lambda_o - \lambda_w$, we obtain an equation for the average pressure and for the water saturation

$$-\nabla \cdot \left(\mathbf{k}\left(\lambda_\Sigma \nabla p + \frac{1}{2}\lambda_\Delta \nabla p_c - (\lambda_w p_w + \lambda_o p_o)\mathbf{g}\right)\right) = \sum_{\alpha=o,w} \frac{F_\alpha}{\rho_\alpha} \tag{7}$$

$$\phi \frac{\partial s_w}{\partial t} - \nabla \cdot \left(\mathbf{k}\lambda_w \left(\nabla\left(p - \frac{1}{2}p_c\right) - \rho_0 \mathbf{g}\right)\right) = \frac{F_w}{\rho_w} \tag{8}$$

We use the following equation for the momentum balance for the oil-water interface (Niessner and Hassanizadeh 2008a)

$$\frac{\partial a_{ow}}{\partial t} - \nabla \cdot (a_{ow} \boldsymbol{k}_{ow} \nabla a_{ow}) = E_{ow} \qquad (9)$$

where $a_{ow}$ is the oil-water interfacial area, $\boldsymbol{k}_{ow}$ is the interfacial permeability and $E_{ow}$ is the rate of production/destruction of interfacial area. Based on physical arguments, Niessner and Hassanizadeh (2008a) proposed the following relation for the rate of production/destruction of interfacial area

$$E_{ow} = \left(\frac{\partial a_{ow}}{\partial p_c}\left(\frac{\partial p_c}{\partial s_w}\right)_{line} + \frac{\partial a_{ow}}{\partial s_w}\right)\frac{\partial s_w}{\partial t} \qquad (10)$$

where the path $(\partial p_c/\partial s_w)_{line}$ is in general unknown, but in the main drainage and imbibition, $p_c$ is a known function of $s_w$. In addition, it is possible to compute the derivative for $e_{wn} = 0$. For all other paths, we interpolate using these three values of $e_{wn} = 0$ (Niessner and Hassanizadeh 2008).

For describing the evolution of the $\beta$ concentration ($\beta = b$ for the bacteria, $\beta = n$ for the nutrients and $\beta = s$ for the surfactants), we consider the following transport equation (Kim 2006, Li et al. 2011)

$$\frac{\partial(\phi s_w C_\beta)}{\partial t} - \nabla \cdot \left(\boldsymbol{D}_\beta \phi s_w \nabla C_\beta - \boldsymbol{u}_w C_\beta - \delta_{b\beta} \boldsymbol{v}_g \phi C_\beta\right) = R_\beta \qquad (11)$$

and the dispersion coefficients are given by

$$D_{\beta,ij} = \delta_{ij} \alpha_{\beta,T} |\boldsymbol{u}| + (\alpha_{\beta,L} - \alpha_{\beta,T})\frac{u_i u_j}{|\boldsymbol{u}|} + \delta_{ij} D_\beta^{eff} \qquad (12)$$

where $C_\beta$ is the concentration, $\boldsymbol{u}_w$ is the water flux, $\delta_{ij}$ is the Kronecker delta, $\boldsymbol{v}_g$ is the settling velocity of bacteria, is the reaction term and $\boldsymbol{u} = \boldsymbol{u}_w/\phi s_w$ is the fluid velocity of the aqueous phase. We consider the following reaction rate terms (Li et al. 2011)

$$R_b = g_{1\,max} \frac{C_n}{K_{b/n}+C_n} \phi s_w C_b - d_1 \phi s_w C_b - \frac{R_s}{Y_{s/b}} \qquad (13)$$

$$R_n = -Y_n \phi s_w C_b - \frac{R_s}{Y_{s/n}} \qquad (14)$$

$$R_s = \mu_{s\,max} \frac{C_n - C_n^*}{K_{s/n}+C_n-C_n^*} \phi s_w C_b \qquad (15)$$

where $g_{1\,max}$ is the maximum bacterial growth rate coefficient, $K_{b/n}$ and $K_{s/n}$ are the half-saturation constants for concentration of specific growth rate and surfactant production respectively, $d_1$ is the bacterial decay rate coefficient, $Y_{s/b}$ and $Y_{s/n}$ are the surfactant yield coefficients per unit bacteria and nutrient respectively, $Y_n$ is the maintenance energy + bacterial growth yield coefficient representing nutrient consumed, $\mu_{s\,max}$ is the maximum specific surfactant production rate and $C_n^*$ is the critical nutrient concentration for metabolism. The reaction term for the bacteria includes Monod-type growing of bacteria, lineal dying of bacteria and a reduction of the bacterial concentration in order to produce the surfactants. The reaction term for the nutrients includes reduction of the concentration in order to produce bacteria and surfactants. The last reaction term includes the production of surfactants.

To model that surfactants are produce on the oil-water interface (Donaldson et al. 1989), we propose the following expression for the specific surfactant production rate

$$\mu_s(C_n, a_{ow}) = \mu_{s\,max} \frac{a_{ow}}{K_a + a_{ow}} \frac{C_n - C_n^*}{K_{s/n}+C_n-C_n^*} \qquad (16)$$

where $K_a$ is a half-saturation constant. The previous relation gives no production of surfactants when there is not interfacial area and the production rate tends to $\mu_{s\ max}$ when $a_{ow}$ increases.

For including in the model the oil-water interfacial tension and residual oil saturation reduction, it is common to give such expression as a function of the surfactant concentration. In Li et al. (2011), they included the following relations (Bang and Caudle 1984, Li et al. 2007)

$$\sigma_{ow}^* = \min\left(\sigma_{ow,min}\left(\frac{\sigma_{ow,max}}{\sigma_{ow,min}}\right)^{\left(\frac{C_{ps,max}-\bar{C}_{ps}}{C_{ps,max}-C_{ps,min}}\right)^s}, \sigma_{ow}\right) \tag{17}$$

$$s_{or}^* = \min\left(s_{or}^{min} + (s_{or}^{max} - s_{or}^{min})[1 + (T_1 N_T)^{T_2}]^{1/T_2 - 1}, s_{or}\right) \tag{18}$$

where $\sigma_{ow}$, $\sigma_{ow,min}$ and $\sigma_{ow,max}$ are the interfacial tension, its minimum and maximum, $C_{ps,min}$ and $C_{ps,max}$ are the surfactant concentration's minimum and maximum, $\bar{C}_{ps}$ is the average surfactant concentration, $s$ is an exponent parameter, $s_{or}^{min}$ and $s_{or}^{max}$ are the residual oil saturation's minimum and maximum, $N_T$ the trapping number and $T_1$ and $T_2$ fitting parameters. From experiments, we have that the interfacial tension reduction influences the capillary pressure and relative permeabilities (Nielsen 2010a). In our model, we use the interfacial tension relation proposed by Nielsen et al. (2010b), we use a linear reduction of capillary pressure (Islam 1990) and the Coats's method (Coats 1980)

$$\sigma_{ow}^* = \sigma_{ow} \frac{-\tanh(l_3 C_s - l_2) + 1 + l_1}{-\tanh(-l_2) + 1 + l_1} \tag{19}$$

$$p_c^* = \sigma_{ow}^* p_c \tag{20}$$

$$f(\sigma_{ow}^*) = \left(\frac{\sigma_{ow}^*}{\sigma_{ow}}\right)^{1/n} \tag{21}$$

$$s_{or}^* = f(\sigma_{ow}^*) \cdot s_{or} \tag{22}$$

$$s_{wi}^* = f(\sigma_{ow}^*) \cdot s_{wi} \tag{23}$$

$$k_{rowi}^* = f(\sigma_{ow}^*) \cdot k_{rowi} + [1 - f(\sigma_{ow}^*)] \cdot 1 \tag{24}$$

$$k_{rwor}^* = f(\sigma_{ow}^*) \cdot k_{rwor} + [1 - f(\sigma_{ow}^*)] \cdot 1 \tag{25}$$

$$a^* = f(\sigma_{ow}^*) \cdot a + [1 - f(\sigma_{ow}^*)] \cdot 1 \tag{26}$$

$$b^* = f(\sigma_{ow}^*) \cdot b + [1 - f(\sigma_{ow}^*)] \cdot 1 \tag{27}$$

where $l_1$, $l_2$, $l_3$ and $n$ are fitting parameters.

In this work, we have performed numerical simulations in 1D. We considered a uniform cell-centered grid with half-cells at the boundaries for the space domain. For the time domain, we considered a uniform discretization form the initial time until the final time. For the space discretisation of the equations, we use a finite volume technique called two-point flux approximation. For the time derivatives, we use backward Euler method. For the transport equations, we approximated the values on the element walls using an upwind update. For the rest of parameters, we compute the average value on the element walls. For solving the pressure, saturation and IFA equations, we use an implicit scheme (Radu et al. 2015, List and Radu 2016). First, we solve the average pressure, water saturation and interfacial area equations iteratively until a convergence criterion is reached. To improve stability, we use a linear approximation of the capillary pressure function (Kou and Sun 2010). After, we solve the three transport equations iteratively until a convergence criterion is reached. Then, we update the interfacial tension, residual oil saturation, irreducible water saturation and parameters for the relative permeabilities. We made the implementation in MATLAB, testing the convergence of the scheme with analytical examples and against Benchmark simulations.

**Numerical experiment**

We consider a porous media of length **L = 1 m**. Table 1 shows the parameters we use in the simulation. We inject on the left side water, bacteria and nutrients, then it results in water, oil, bacteria, nutrients, and surfactants flowing out on the right side. In order to solve the system of equations, it is necessary to set the initial and boundary conditions. Regarding the oil and water pressures, we set $p_o(x,0) = 9.417\ kPa$ and $p_w(x,0) = 0.981\ kPa$ respectively. Then, the initial average pressure is $p(x,0) = 5.199\ kPa$ and the initial capillary pressure is $p_c(x,0) = p_o(x,0) - p_w(x,0) = 8.436\ kPa$. We consider that the initial water saturation is $s_w(x,0) = 0.3$. Given the initial water saturation and capillary pressure, we can compute the initial interfacial area $a_{ow}(x,0) = a_{ow}(s(x,0), p_c(x,0))$. We consider that initially the bacterial, nutrient and surfactant concentration inside the porous media is 0. For the boundary conditions, we consider a flux boundary condition on the left boundary $Q_T/A(x,0) = 2 \times 10^{-5}\ ms^{-1}$, where $Q_T$ is the injection/production rate and $A$ is the cross-sectional area, while we set a constant pressure on the right boundary $p(L,t) = 5.199\ kPa$. Due to we just inject water, then the left boundary condition for the water saturation is $s_w(0,t) = 1 - s_{sor}(0,t)$. We consider a Dirichlet boundary condition for the interfacial area on the left boundary, evaluating with the current water saturation and initial capillary pressure. Zero Neumann boundary conditions are considered for the water saturation and interfacial area on the right boundary. After one hour of water injection, we start to inject bacteria and nutrients on the left boundary at concentrations of $0.5\ kg/m^3$ and $0.3\ kg/m^3$ respectively. For the surfactants on the left boundary, we consider a non-flux boundary condition. We consider zero Neumann boundary conditions for the three concentrations on the right boundary.

*Table 1 List of parameters for the numerical experiment.*

| Parameter | Value | Parameter | Value |
|---|---|---|---|
| $\phi_0$ | $0.4$ | $k_0$ | $1 \times 10^{-12}\ m^2$ |
| $\mu_w$ | $1 \times 10^{-3}\ kg/(m \cdot s)$ | $\mu_o$ | $3.92 \times 10^{-3}\ kg/(m \cdot s)$ |
| $g_{1\ max}$ | $2 \times 10^{-5}\ /s$ | $\mu_{p\ max}$ | $1.5 \times 10^{-5}\ /s$ |
| $g$ | $0$ | $v_g$ | $0$ |
| $\alpha_{b,T}$ | $0.01\ m$ | $D_b^{eff}$ | $1.5 \times 10^{-9}\ m^2/s$ |
| $\alpha_{n,T}$ | $0.01\ m$ | $D_n^{eff}$ | $1.5 \times 10^{-9}\ m^2/s$ |
| $\alpha_{s,T}$ | $0.01\ m$ | $D_s^{eff}$ | $1.5 \times 10^{-9}\ m^2/s$ |
| $K_{b/n}$ | $1\ kg/m^3$ | $K_{p/n}$ | $1\ kg/m^3$ |
| $Y_{s/b}$ | $7.944$ | $Y_{s/n}$ | $1.144$ |
| $l_1$ | $41 \times 10^{-4}$ | $\rho_b$ | $1600\ kg/m^3$ |
| $l_2$ | $0$ | $\rho_o$ | $800\ kg/m^3$ |
| $l_3$ | $500$ | $\rho_w$ | $1000\ kg/m^3$ |
| $s_{wi}$ | $0.2$ | $s_{or}$ | $0.3$ |
| $K_a$ | $1000\ /m$ | $k_{wo}$ | $5 \times 10^{-8}\ m^3/s$ |
| $F_o$ | $0$ | $F_w$ | $0$ |
| $d_1$ | $1 \times 10^{-5}\ /s$ | $Y_n$ | $1.5 \times 10^{-5}\ /s$ |
| $\sigma_0$ | $3.37 \times 10^{-2}\ N/m$ | $C_n^*$ | $0$ |
| $\alpha_1$ | $170.936$ | $\alpha_2$ | $0$ |
| $\alpha_3$ | $1.244$ | $\alpha_4$ | $-0.963$ |
| $n$ | $4$ | | |

**Results and discussion**

We run the code simulating 20 hours of injection. In Figure 1, we observe the spatial distribution of the water saturation at five times. When we start the injection, the oil is recovered due to waterflooding. We observe that after 10 hours of injection, the water saturation is greater than 0.7, this due to the surfactant action. Analysing the average pressure, we observe that the entry pressure it is decreasing over time in order to maintain the same inlet and outlet flux.

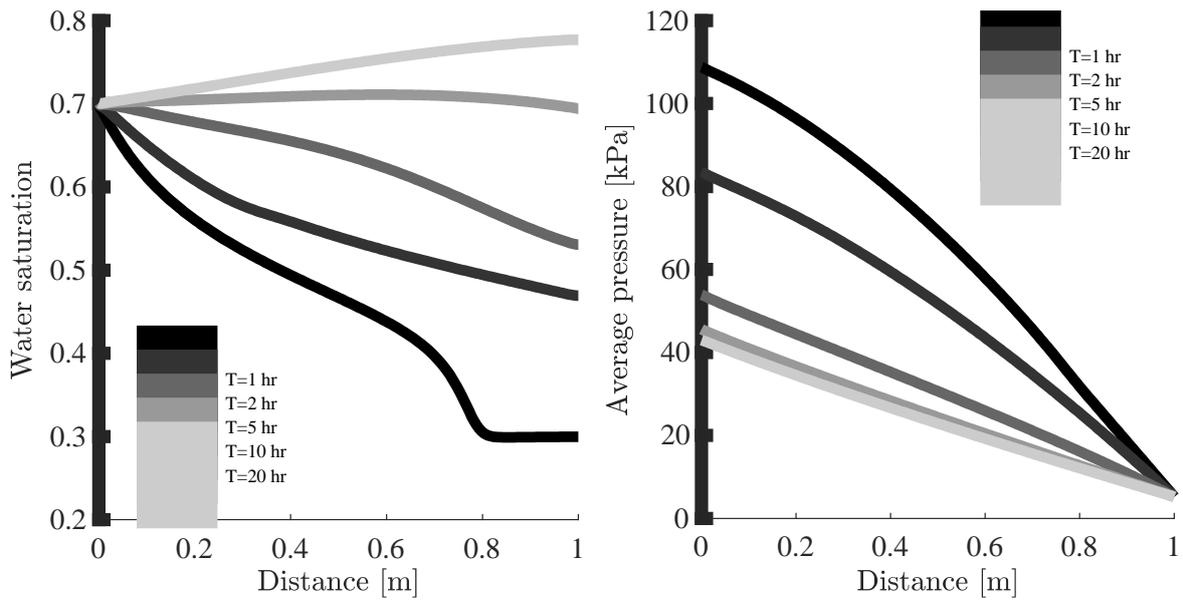

*Figure 1 Water saturation and average pressure profiles at different times of injection.*

Figure 2 shows the interfacial area and capillary pressure profiles at five different times of injection. We observe that the interfacial area decreases over time, due to the porous media is being filled with water. The second graph shows the capillary pressure dynamics over time.

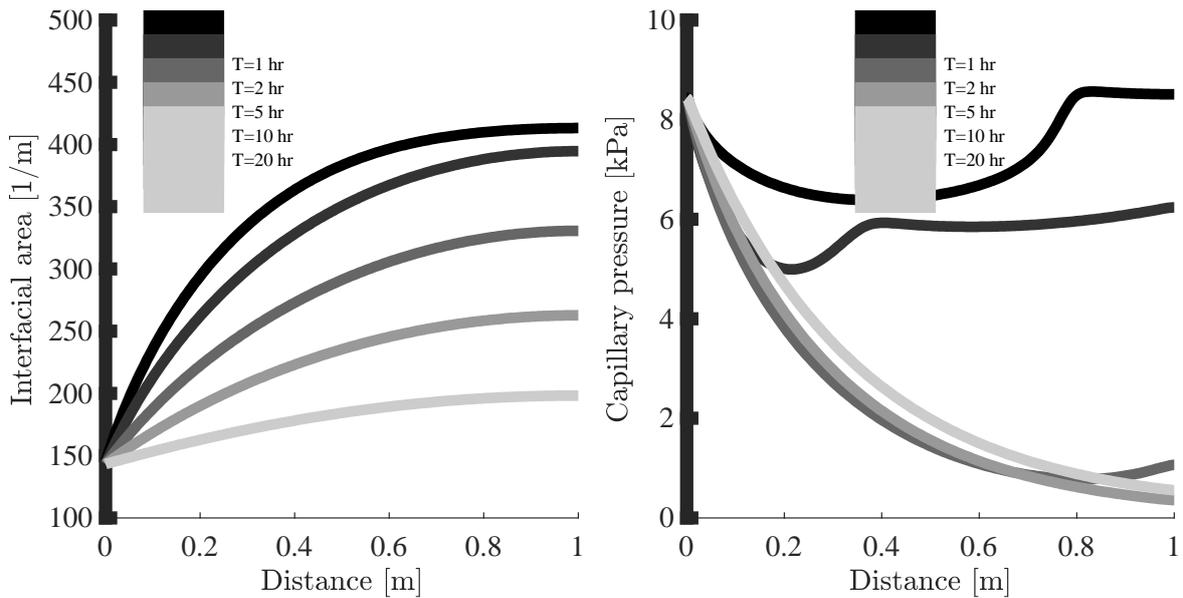

*Figure 2 Interfacial area and capillary pressure profiles at different times of injection.*

Figure 3 shows the spatial distribution of bacteria, nutrient and surfactant at three different times. We observe that some nutrients are consumed in order to produce more bacteria and surfactants. Moreover, we observe that the bacterial concentration is increasing over time due to we are continuously injecting nutrients and we did not include the bacteria absorption in the model.

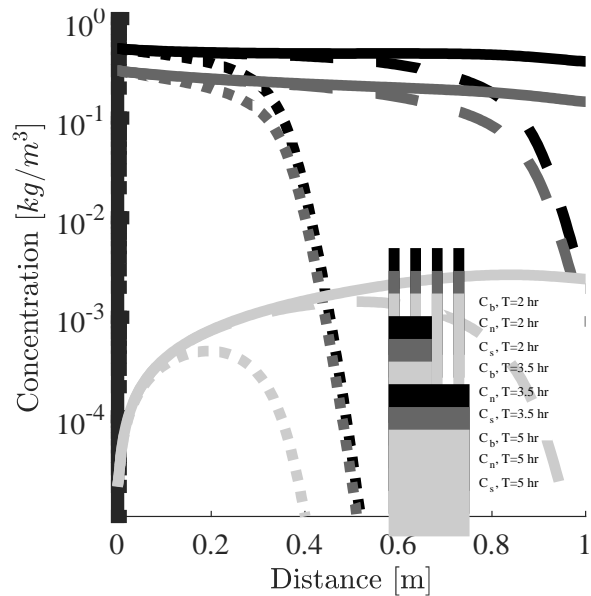

*Figure 3 Bacterial, nutrient and surfactant concentration profiles at different times of injection.*

To study the impact of the inclusion of surfactant production on the oil-water interface, Figure 4 shows the residual oil saturation profiles at 5 different times of injection for two different simulations: the first graph corresponds to the MEOR model not including the surfactants production on the oil-water interface, while the second graph includes it. We observe less residual oil recovery including the surfactant production on the oil-water interfacial area. From Eq. 16, we observe that the production rate is less when we include the interfacial area. Then, the surfactant production is overestimated, leading to a greater oil recovery when we do not include the interfacial area. We also noticed that most of the residual oil near to the injection boundary it is not recovered. This is because we inject bacteria and nutrients to the porous media, so the surfactant production is performed inside the reservoir and at the beginning the surfactant concentration is not high enough to decrease significantly the interfacial tension. One strategy for recovering this oil is to stop the production and keep injecting nutrients, so the bacteria will growth and produce more surfactants near the inlet boundary.

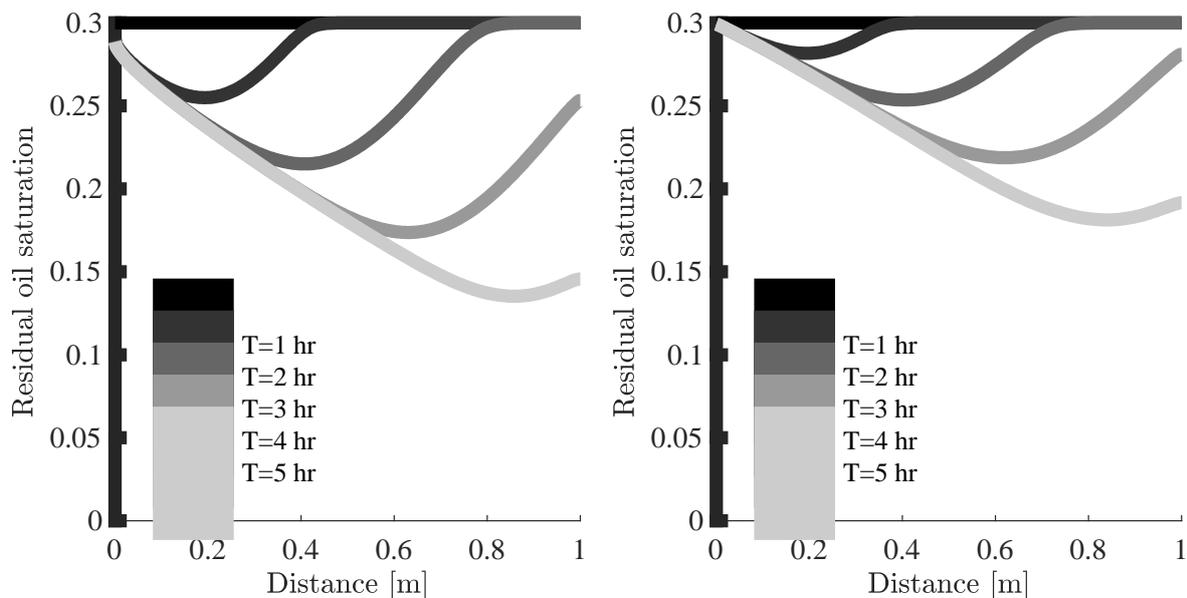

*Figure 4 Residual oil saturation profiles at different times of injection. The left graph does not include the surfactant production on the IFA.*

Finally, Figure 5 shows the oil recovery as a function of the injected pore volume. For this example, we observe that after injection 0.75 pore volume, the surfactant concentration it is enough for recovering the trapping oil. As we explained previously, we also observe less oil recovery when we consider the surfactant production on the oil-water interface.

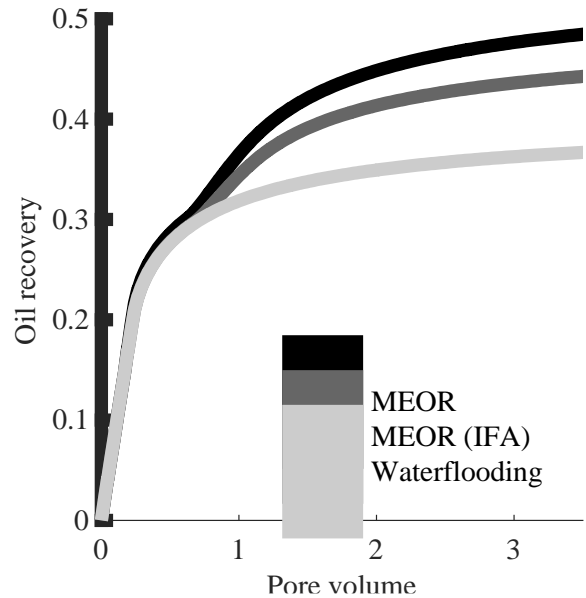

*Figure 5* *Comparison of the oil recovery due to the surfactant action.*

**Conclusions**

We extend a previous core-scale MEOR model by incorporating the oil-water interfacial area into it. This model includes two-phase flow, bacteria, nutrient and surfactant transport and considers the interfacial area and reduction of residual oil saturation and irreducible water saturation due to the action of surfactants. To our knowledge, this is the first time that the role of the oil-water interfacial area in MEOR is studied. Inclusion of the interfacial area reduces hysteresis and allows to model new important mechanisms. In this work, we proposed a model including the biological production of surfactants at the oil-water interface. The numerical simulations show the dynamics involved on the principal mechanism in MEOR. We obtained less oil recovery when we included the biological production of surfactants at the oil-water interface, due to the specific surfactant production rate is overestimated when we do not include it. It is necessary to perform laboratory experiments in order to calibrate the parameters and validate the model assumptions. Moreover, there are important processes that should be included in the model; for example, absorption of nutrients, bacteria and surfactants in the solid matrix, porosity and permeability changes due to biomass attachment to the solid matrix and transport of surfactants at the oil phase. In addition, it is necessary to investigate new relationships for the rate of production/destruction of interfacial area (we just found one relation in the literature).

**Acknowledgments**

We acknowledge financial support for the IMMENS project from the Research Council of Norway (project no. 255426).